\documentclass[preprint,aps,amssymb,showpacs,superscriptaddress,nofootinbib]{revtex4}
\usepackage{graphicx}

\usepackage{latexsym}
\usepackage{amssymb, amsmath}
\usepackage{graphicx}
\usepackage{dcolumn}
\usepackage{bm}
\def\ee{\end{equation}}
\def\beq{\begin{eqnarray}}
\def\eeq{\end{eqnarray}}
\def\bn{\begin{eqnarray*}}
\def\en{\end{eqnarray*}}
\def\w{\omega}
\def\g{\gamma}
\def\S{\Sigma}

\def\a{\alpha}
\def\b{\beta}
\def\m{\mu}
\def\n{\nu}
\def\D{\Delta}
\def\d{\delta}

\def\l{\lambda}
\def\th{\theta}

\def\e{\epsilon}

\def\he{\hat =}

\def\p{\phi}
\def\G{\Gamma}

\begin{document}

\title{Schwarzschild horizon dynamics and $SU(2)$ Chern-Simons theory}

\author{Romesh K. Kaul}
\email{kaul@imsc.res.in}
\affiliation{The Institute of Mathematical Sciences, CIT Campus, Chennai-600 113, INDIA.}
 
\author{Parthasarathi Majumdar }
\email{parthasarathi.majumdar@saha.ac.in}
\affiliation{Saha Institute of Nuclear Physics, AF/1 Bidhannagar, Kolkata 700 064, INDIA.}
 
\begin{abstract}
We discuss the effect of different choices in partial gauge fixing of bulk local
Lorentz invariance, on the
description of the horizon degrees of freedom of a Schwarzschild
black hole as an $SU(2)$ Chern-Simons theory with specific sources. A classically
equivalent description in terms of an $ISO(2)$ Chern-Simons
theory is also discussed. Further, we demonstrate that both these descriptions can be partially gauge
fixed to a horizon theory with $U(1)$ local gauge invariance, with the solder
form sources being subject to extra constraints in directions orthogonal to an
internal vector field left invariant by $U(1)$ transformations. Seemingly
disparate approaches on characterization of the horizon theory for the
Schwarzschild black hole (as well as spherical Isolated Horizons in general) are thus shown to
be equivalent physically.    
\end{abstract}


\maketitle
 
 
\section{Introduction}

The event horizon (EH) of a black hole spacetime (and more generally an Isolated
Horizon (IH)) \cite{ash1}), is a null inner boundary of the part 
of the entire spacetime manifold accessible to asymptotic observers. It has
the topology of $R \times S^2_{}$ and a {\it degenerate} intrinsic
three-metric. Because of this latter property, it is not possible to describe
the horizon degrees of freedom in terms of a Lagrange density with standard
kinetic terms where contractions are usually made with the inverse metric. In
this sense, the horizon three-fold does not support any local propagating
degree of freedom. The only possible degrees of
freedom on the horizon have to be global or topological, described by a {\it
topological} (metric independent) quantum field theory. Three dimensional Chern-Simons theories
appear to be good candidate topological field theories for this description.
  
In Loop Quantum Gravity (LQG), bulk spacetime properties  
are described in terms of the Barbero-Immirzi
family of $SU(2)$ connections \cite{rovbk} obtained from a partially gauged
fixed $SO(1,3)$ theory. All physics associated with bulk
spacetime geometry must be invariant under local $SU(2)$ transformations.  
Since, at the classical level, the degrees of
freedom and their dynamics on an EH (IH) are {\it completely determined by the
  geometry and dynamics in the bulk}, the theory of the horizon degrees of freedom,
has to imbibe this $SU(2)$ gauge invariance from the bulk. This implies that
the horizon degrees of freedom should  be described by
a topological $SU(2)$ Chern-Simons theory on the three-manifold
$R\times S^2_{}$, coupled to appropriate sources  derived
from tetrad components in the bulk. However, there are ambiguities in
partially gauge fixing the bulk local Lorentz invariance to $SU(2)$. Studying the
effect of these on the horizon theory is the main thrust of this paper. 
 
Use of $SU(2)$ gauge theory to count the microstates associated with a
two-dimensional surface has a long history. Inspired by the proposal of Crane
that quantum gravity be described  by a topological field theory \cite{crane}
and the holographic hypothesis of 't Hooft and Susskind \cite{hooft}, it was
Smolin who first explored the use of $SU(2)$ Chern-Simons theory induced on
boundary satisfying self-dual boundary conditions in Euclidean gravity and
also demonstrated that such a boundary theory obeys the Bekenstein bound
\cite{smo}. This was followed by the work of Krasnov who applied  these ideas
to the black hole horizon and used the ensemble of quantum states of  $SU(2)$
Chern-Simons theory associated with  the spin assignments of the punctures on
the surface to count the microstates,   leading to an area law for the entropy
\cite{kras}. The coupling of the Chern-Simons theory was argued to be
proportional to the horizon area and also inversely proportional to the
Barbero-Immirzi parameter $\g$.  This was the first application of $SU(2)$
Chern-Simons theory for calculating the black hole entropy.  On the other
hand, within the Loop Quantum Gravity, assuming that the geometry of the
fluctuating  black hole horizon is given by the quantum states associated with
the intersections of knots carrying $SU(2)$ spins impinging on  the
two-dimensional surface,  a counting procedure was developed by Rovelli,
again obtaining an area law for the entropy \cite{rov}. In the general context
of Isolated Horizons, application of $SU(2)$ Chern-Simons theory as a
boundary theory came with the work of Ashtekar, Baez, Corichi and Krasnov
\cite{ash2} and was further developed in ref.s. \cite{kmplb, kmprl, dkmprd,
krprd}.

Following the derivation of the area law for the entropy of large area IHs in
\cite{ash2}, corrections due to quantum spacetime fluctuations, leading
logarithmic in area $-{\frac 3 2} \ln A$ (with this definite coefficient
$-3/2$) and subleading in inverse powers of area, were obtained within the framework of this
$SU(2)$ Chern-Simons theory in \cite{kmprl}. These were
done in the approximation where spin $1/2$ representations were placed on the punctures of the 
spatial slice $S^2_{}$ of the horizon. Such configurations provide the dominant
contribution to the dimensionality of the IH Hilbert space. The coefficient of the leading area term
depends on the Barbero-Immirzi parameter $\g$. Matching this with
the Bekenstein-Hawking area law fixes a definite value of $\g$. In fact, the
logarithmic corrections to the area law, obtained in this framework, are the first ever 
{\it signature} corrections thrown up by quantization of IHs within LQG, obtained by {\it  using  
Chern-Simons theories}. That these logarithmic corrections do not depend on the 
value of $\g$ also emerges from these studies. An improvement over the
approximation used in these  calculations has been achieved by  including
the contributions of spins other than $1/2$ on some of the punctures \cite{meiss}, 
which changes the coefficient of the leading area term and thus improves the value of 
$\g$  by about 10\%. In these  counting schemes, however, the logarithmic correction,  
$-{\frac 3 2} \ln A$, which does not depend on $\g$, is unaffected. 
In fact this leading log(area) correction is rather generally insensitive to the 
value of the spins placed on the punctures. For example, it has been explicitly
shown that placing  spin 1 representations on all the punctures changes the value of $\g$,  
but leaves the coefficient of the leading logarithmic correction unchanged \cite{krprd}.

Recently, there has been a resurgent interest in this  $SU(2)$ Chern-Simons theoretic description 
of Isolated Horizons started by \cite{per} and followed by others \cite{kr, agui, sahl, cm}.
Some of these papers have recalculated and confirmed the nature of the
leading logarithmic correction to the Bekenstein-Hawking area law  for
microcanonical entropy of isolated horizons, with the
definite coefficient $-3/2$, found a decade earlier in \cite{kmprl,
dkmprd}. However, in these latter formulations, the coupling strength of the
Chern Simons mysteriously appears to diverge for a value of the Barbero-Immirzi
parameter which seems to have no particular significance.   

The $SU(2)$ Chern-Simons description of the horizon degrees of freedom has occasionally
been viewed in the literature as a {\it counterpoint} to
the description in terms of a $U(1)$ Chern-Simons theory \cite{ash1, ash2}.  
These, apparently disparate, points of view, are in fact quite reconcilable. 
The result follows from the fact that what is relevant in the problem on hand are properties of 
fields on the spatial slice $S^2_{}$ of the horizon. It is indeed always possible to 
partially gauge 
fix the   $SU(2)$  theory on  $S^2$ to a  theory with only a left over 
$U(1)$ invariance.   In particular, as argued in \cite{bkm},  to go from the $SU(2)$ theory to the $U(1)$ theory  in the
gauge fixed formulation, there are additional constraints for 
the solder forms orthogonal to the direction specified by an internal space unit vector
left invariant by a $U(1)$ subgroup of $SU(2)$ gauge group. These constraints
only reflect the $SU(2)$ underpinnings of the $U(1)$ theory. 

This special property of being able to fix an $SU(2)$ gauge invariance to a $U(1)$ gauge invariance,
obtains only on $S^2$. 
One direct way of  seeing  this is  as follows: In an $SU(2)$ gauge  theory described
through the triplet of field strength $F^{(i)}_{\theta \phi}  ~(i=1,2,3) $  on
$S^2$, we can always rotate the field strength  through an $SU(2)$ gauge
transformation to have only one nonzero component lying in the direction
preserved by a subgroup $U(1)$: $F^{(i)}_{\theta \phi} 
\rightarrow F'^{(i)}_{\theta \phi} = (
F'^{(1)}_{\theta \phi}~, ~0~,~0)$. Next,  for such a field strength on $S^2_{}$,
the antisymmetric two-tensor $F'^{(1)}_{\theta \phi}$ 
is given by the curl of a  vector field with components $A_{\theta}, A_{\phi}$:
$F'^{(1)}_{\theta \phi } = \partial_{\theta} A_\phi - \partial _\phi
A_\theta$, which defines the $U(1)$ curvature.
Clearly, this gauge fixing implies that $SU(2)$ and $U(1)$ gauge fields on $S^2$ have
the {\it same physical content.}  
 
In the present paper, we revisit the $SU(2)$ Chern-Simons
description of the Schwarzschild event horizon, and discuss effects on the
horizon induced by various ways of partial gauge fixing of bulk local Lorentz
invariance. In particular, the mysteriously diverging Chern-Simons coupling 
found recently in \cite{per} is seen to emerge straightforwardly as a
consequence of this gauge fixing. An equivalent description in terms of an 
$ISO(2)$ Chern-Simons theory is also
discussed in this context. How an effective Chern-Simons
theory, with these higher gauge invariances all gauge fixed to a $U(1)$, ties
up these approaches, with appropriate constraints corresponding to
the gauge fixing, is explained in some detail. These results emerge within a discussion of 
properties of the future EH of the Kruskal-Szekeres extension of the
Schwarzschild spacetime, but generalize to any spherical Isolated
Horizon. Dealing with an exact black hole solution allows
us to extract information on its horizon dynamics in a manner that is physically
equivalent to extant approaches based on the Hamiltonian analysis of isolated
horizons \cite{ash1}, at least for this particular solution. Is this generic enough for
all spherical isolated horizons? In what follows, we point out the precise
features in our results which also emerge in the general case of spherical isolated horizons
obtained in earlier work cited above. 

In Section II, we display an appropriate set of the tetrad components and corresponding spin
connection components. We explicitly exhibit the gauge equivalent class of tetrads
in terms of a single function $\alpha(x)$ which is the sole ambiguity in the
choice of a local {\it Lorentz frame} for the Schwarzschild metric. Other tetrad sets are related to ours
only through different choices of {\it coordinates}. On the black hole
(future) horizon, in Section III, the field strength 
components associated with our connection fields are shown to satisfy a set of equations   
which exhibit a left over invariance under $U(1)$ gauge transformations, when
the scale function assumes certain specific values. We shall then further
demonstrate that, for other local Lorentz frame choices in the bulk, these
horizon equations can also be interpreted as a gauge fixed version of  
an $ISO(2)$ Chern-Simons theory. In addition, there is, for different local
Lorentz frame choices,  an alternative description in terms of a  
Chern-Simons theory of the Barbero-Immirzi $SU(2)$ gauge fields which in a gauge fixed 
version reproduces the  $U(1)$ gauge theory. This will be presented in Section IV.
In this gauge fixed $U(1)$ formulation, the sources in the direction
orthogonal to $U(1)$  subgroup are constrained to vanish. In particular, as
emphasized in the  earlier analysis in \cite{bkm}, 
the two components of $SU(2)$ triplet solder forms on the spatial slice of the horizon
orthogonal to the direction specified by the $U(1)$ subgroup   are indeed 
zero as they should be.  Finally, Section V will contain a few concluding remarks.
  
While our analysis presented here is for the future (black hole) horizon of the 
Kruskal-Szekeres extended  Schwarzschild spacetime, rather than the
past (white hole) horizon, similar conclusions would ensue for that case as well.

\section{Schwarzschild metric in Kruskal-Szekeres coordinates}

  The Schwarzschild metric, expressed in  the Kruskal-Szekeres null coordinates $v$ and $w$, is:
\beq
ds^2_{}= -2 A(r)~ dv dw ~+~ r^2 (v,w) ~ \left( d \theta^2_{}  +  \sin^2_{} \theta d \p^2_{}\right),  
 ~~~~A(r) = {\frac {4 r^3_0} r} \exp \left( -  {\frac r {r^{}_0}}\right) \label{metricKruskal}
\eeq
where $r$ is given implicitly  by:
\beq
- ~2 v w~=~  \left( {\frac r {r^{}_0}} - 1\right) ~\exp \left( {{\frac r {r^{}_0}}}_{}\right)
\eeq
The exterior region ($r> r^{}_0$)  of the black hole is given by:  
$ v w < 0 ~,  ~ v>0~, ~w<0$. The interior region ($ r^{}_0>r>0$) is:  
  $ 0< 2 vw <  1~, ~v>0~,~ w>0$.
In terms of these coordinates, the past and future event horizons are given by:
$v w~=~ 0$. The outgoing null geodesics are given by $w ~= ~constant$ and the
ingoing null geodesics  by $v~=~ constant$. The curvature singularity ($r=0$)
is described by $2 v w = 1$.

Corresponding to the metric (\ref{metricKruskal}), non-zero Christoffel symbols $\Gamma^{~~\l}_{\m\n} $  are: 
\beq
&& \G^{~~~v}_{vv}~=~ \partial^{}_v \ln A~, ~~~~ \G^{~~~v}_{\theta \theta}~=~ {\frac {r^2_{}} A} ~\partial^{}_w \ln r~, 
~~~~ \G^{~~~v}_{\p\p}~=~ {\frac {r^2_{} \sin^2_{} \theta} A}~\partial^{}_w \ln r ~; \nonumber \\
&&\G^{~~~w}_{ww}~=~ \partial^{}_w \ln A~, ~~~ \G^{~~~w}_{\theta \theta}~=~ {\frac {r^2_{}} A} ~\partial^{}_v \ln r~, 
~~~~ \G^{~~~w}_{\p\p}~=~ {\frac {r^2_{} \sin^2_{} \theta} A} ~\partial^{}_v \ln r ~; \nonumber \\
&& \G^{~~~\theta}_{v \theta}~=~ \partial^{}_v \ln r~, ~~~~~ \G^{~~~\theta}_{w \theta} ~=~ \partial^{}_w \ln r~, ~~~ ~~~~
\G^{~~~\theta}_{\p \p} ~=~ - ~\sin \theta \cos \theta~;  \nonumber \\
&& \G^{~~~\p}_{v \p}~=~ \partial^{}_v \ln r~, ~~~~~ \G^{~~~\p}_{w \p} ~=~ \partial^{}_w \ln r~, ~~~~~~~ 
\G^{~~~\p}_{\theta \p} ~=~ \cot \theta
\eeq 

Now we choose an appropriate set of tetrad fields which are compatible with the
metric (\ref{metricKruskal}). In the following, we shall restrict ourselves only to the exterior region 
of the black hole $(v>0~,~ w<0)$. In this region, we take the tetrad fields as:
 \beq
e^0_\m &=&  
  {\sqrt {\frac A  2}}\left( {\frac w {\a}} ~\partial_\m^{}v +{\frac {\a} w} ~\partial^{}_\m w \right)~,  ~~~~~
 e^1_\m ~=~    {\sqrt {\frac A  2}}\left({\frac  w {\a}} ~\partial_\m^{}v -{\frac {\a}w} ~\partial^{}_\m w \right)~, \nonumber \\
e^2_\m & =&   r ~\partial^{}_\m \th~, ~~~~~~~~~~~~~~~
e^3_\m ~= ~ r\sin \th ~\partial^{}_\m \p \label{tetradsexKruskal}
\eeq

Here $\a$ is an arbitrary function of the coordinates; every choice of $\a(x)$
characterizes the local Lorentz frame in the indefinite metric plane ${\cal I}$ of the Schwarzschild spacetime
whose spherical symmetry implies that it has the topology ${\cal I} \otimes S^2$. Corresponding to these tetrad
fields, the spin connections satisfying the relation $\partial^{}_\m e^I_\n - \G^{~~\l}_{\m\n} ~e^I_\l + \w^{~I}_{\m~J}
~e^J_\n = 0$  are
 \beq
\w^{~01}_\m &=& -{\frac 1 2} \left( 1-{\frac {r^2_0} {r^2_{}}}\right) {\frac 1 v}~   \partial^{}_\m v ~-~
{\frac 1 2} \left( 1+ {\frac {r^2_0} {r^2_{}}} \right) {\frac 1 w} ~\partial_\m w + \partial_\m \ln \a~,
~~~~~ \w^{~23}_\m = -~\cos \theta ~ \partial^{}_\m \p  \nonumber \\
 \w^{~02}_\m &=& - {\sqrt {\frac A 2}} ~ {\frac 1 {2r^{}_0}} \left( {\frac
     {vw} {\a}}  + \a \right) \partial_\m \th ~,  ~ ~~~~~~~\w^{03}_\m ~=~  -
 {\sqrt {\frac A 2}} ~ {\frac {\sin \th} {2r^{}_0}} \left( {\frac {vw} {\a}} +
  \a \right) \partial_\m \p ~  \nonumber \\
\w^{~12}_\m &=&   - {\sqrt {\frac A 2}} ~ {\frac 1 {2r^{}_0}}   \left( {\frac
    {vw} {\a}} - \a \right) \partial_\m \th ~,  ~ ~~~~~~~ 
\w^{~13}_\m=   - {\sqrt {\frac A 2}} ~ {\frac {\sin \th} {2r^{}_0}} \left(
  {\frac {vw} {\a}} - \a \right) \partial_\m \p ~  \label{spinconnKruskal}
\eeq
The curvature tensor $R^{~~~IJ}_{\m\n} = \partial^{}_\m \w^{~IJ}_\n - \partial^{}_\n \w^{~IJ}_\m
+\w^{~IK}_\m \w^{~~~J}_{\n K} - \w^{~IK}_\n \w^{~~~J}_{\m K}$ for the spin connections (\ref{spinconnKruskal}) is given by:
\beq
&& R^{~~~01}_{\m\n}~=~ {\frac {2 r^{}_0} {r^3_{}}} ~ \S^{01}_{\m\n}~,
~~~~~R^{~~~02}_{\m\n} ~=~-~ {\frac {r^{}_0} {r^3_{}}}~ \S^{02}_{\m\n}~,
~~~~ ~R^{~~~03}_{\m\n} ~=~- ~{\frac {r^{}_0} {r^3_{}}}~ \S^{03}_{\m\n} ~, 
\nonumber \\
&&R^{~~~23}_{\m\n} ~=~ {\frac {2 r^{}_0} {r^3_{}}}~\S^{23}_{\m\n}~, 
~~~~~ R^{~~~31}_{\m\n} ~=~- ~{\frac {r^{}_0} {r^3_{}}}~ \S^{31}_{\m\n} ~,
~~~~~ R^{~~~12}_{\m\n} ~=~- ~{\frac {r^{}_0} {r^3_{}}}~ \S^{12}_{\m\n} \label{RKruskal}
\eeq
where the solder forms $\S^{IJ}_{\m\n} ~= ~e^I_{[\m}~ e^J_{\n]} ~\equiv~
{\frac 1 2} \left( e^I_\m ~e^J_\n - e^I_\n ~e^J_\m \right) $, in the
exterior region ($v>0~,~ w<0$), are
\beq
\S^{01}_{\m\n}&=& -~A ~\partial_{[\m} v~ \partial^{}_{\n]} w
~,~~~~~~\S^{23}_{\m\n}~=~ r^2_{} \sin \theta ~\partial^{}_{[\m} \theta ~ \partial^{}_{\n]} \p~~~~ \nonumber  \\
\S^{02}_{\m\n}&=& r ~{\sqrt {\frac A 2}} ~\left( { \frac w {\a}}
~\partial_{[\m} v ~\partial_{\n]} \theta  ~
+~ {\frac {\a} w} ~\partial_{[\m} w~ \partial_{\n]} \theta \right), \nonumber \\
\S^{03}_{\m\n}&=& r \sin \theta  ~  {\sqrt {\frac A 2}} ~\left( { \frac w {\a}}
~\partial_{[\m} v ~\partial_{\n]} \p  ~
+~ {\frac {\a} w} ~\partial_{[\m} w~ \partial_{\n]} \p \right)~, \nonumber \\
\S^{12}_{\m\n}&=& r ~{\sqrt {\frac A 2}} ~\left( { \frac w {\a}}
~\partial_{[\m} v ~\partial_{\n]} \theta  ~
-~ {\frac {\a} w} ~\partial_{[\m} w~ \partial_{\n]} \theta \right)  \nonumber \\
\S^{31}_{\m\n}&=& -~  r \sin \theta ~   {\sqrt {\frac A 2}} ~\left( { \frac w {\a}}
~\partial_{[\m} v ~\partial_{\n]} \p  ~
-~ {\frac {\a} w} ~\partial_{[\m} w~ \partial_{\n]} \p \right)
   \label{SigmaKruskal}
\eeq

Next,
  LQG is described in terms of Barbero-Immirzi  $SU(2)$ gauge  fields \cite{rovbk}
which are 
linear combinations of the the connection components involving the Barbero-Immirzi parameter
 $\g$. To make contact with this, we introduce the $SU(2)$ gauge field:
\beq
A^{ (i)}_\m~=~   \g \w^{0i}_\m - {\frac 1 2}\e^{ijk}_{} \w^{jk}_\m  \label{gaugeA} 
\eeq  
Substituting for the spin connections from (\ref{spinconnKruskal}), this yields in the exterior region 
($v>0~,~ w<0$):
\beq
A^{(1)}_\m &=& \g ~ \w^{~01}_\m - \w^{~23}_\m = -~ {\frac {\g} 2} \left[
 \left( 1- {\frac {r^2_0} {r^2_{}}} \right) {\frac 1 v} ~\partial_\m v + 
 \left( 1+ {\frac {r^2_0} {r^2_{}}}\right) {\frac 1 w} ~ \partial_\m w  -2 \partial_\m \ln\a \right] 
 + \cos \th~ \partial_\m \p \nonumber \\
 A^{(2)}_\m &=& \g ~ \w^{~02}_\m - \w^{~31}_\m =  - ~{\sqrt {\frac A 2}}~ {\frac 1 {2r_0}} \left[ \g \left( {\frac {vw} {\a}} + \a \right) \partial_\m \th + \sin \th~ \left( {\frac {vw} {\a}} - \a \right)
 \partial_\m \p \right] ~, \nonumber \\
 A^{(3)}_\m &=& \g ~ \w^{~03}_\m - \w^{~12}_\m =  - ~{\sqrt {\frac A 2}}~ {\frac 1 {2r_0}} \left[ \g \sin \th ~\left( {\frac {vw} {\a}} + \a \right) \partial_\m \p - \left( {\frac {vw} {\a}} - \a \right)
 \partial_\m \th \right] \label{BIA}
  \eeq 
   
The choice of the tetrad fields as in eqn.(\ref{tetradsexKruskal}) is not unique; we could have used any
other choice compatible with the metric (\ref{metricKruskal}). 
  
Now let us restrict our discussion  to the event horizon   by taking the limit to the
horizon from the exterior region to unravel the properties of the various fields on the horizon.

 \section{Black hole  horizon and $ISO(2)$ Chern-Simons theory} 

The black hole horizon is the future horizon   given by $w  =0$,
which is a null three-manifold $\D$, topologically $ R\times S^2_{}$, spanned
by the coordinates $a= (v ~, ~\theta~,~ \p)$ where $ 0 <  v < \infty ~, ~0\le
\theta < \pi~,~ 0\le \p<2\pi$. The evolution parameter is $v$.  The null and
future directed geodesics given by $v = constant$  are infalling into $\D$.
The foliation of the manifold $\D$ is provided by $v=constant$ surfaces,  each an  $S^2_{}$.

The relevant tetrad fields $e^I_a$ from (\ref{tetradsexKruskal})  on $\D$ are given by:
$e^0_a ~\he~0~, ~~ e^1_a ~\he ~ 0~, $ $ e^2_a ~\he~ r^{}_0 ~\partial^{}_a \th~, $ $ ~e^3_a ~\he~ r^{}_0 \sin \th ~ \partial^{}_a \p$ where $a= (v~, ~ \th~ ,~ \p)$ (we denote equalities on $\D$, that is for $w=0$, 
 by the symbol $ ~\he~$). The intrinsic metric on $\D$ is: $q^{}_{ab} = e^I_a e^{}_{Ib} $ $~\he~ m^{}_a {\bar m}_b
 + m^{}_b {\bar m}^{}_a$ with $m^{}_a ~\equiv~ {\frac {r^{}_0} {\sqrt 2}} \left( \partial^{}_a \th+ i \sin \th ~ \partial^{}_a \p \right) $. This metric is indeed degenerate with its signature $(0~,~+~,~+~)$.
 
  Notice that the solder fields  on the horizon $\D$ 
    are: 
  \beq
 {\rm all} ~~~~\S^{IJ}_{ab} ~~\he~~ 0   ~~~~~ {\rm except} ~~~  \S^{23}_{ab} ~~\he~~ r^2_0 \sin \th~ 
   \partial_{[a} \th ~ \partial^{}_{b]} \p
  \label{S}
  \eeq
  and the spin connection fields are:
    \beq
 \w^{~01}_a ~&\he&~ {\frac 1 2}~\partial_a\ln \b~, ~~~~~~~~~ 
 \w^{~23}_a ~=~ -\cos \th~ \partial_a \p~, \nonumber \\
  \w^{~02}_a ~&\he&~  -~      {\sqrt \b }~ \partial_a \th~, 
 ~~~~~ \w^{~03}_a ~~\he~~ -~ {\sqrt {\b}}~ \sin \th  ~ \partial_a \p  \nonumber \\
 \w^{~12}_a ~& \he &~      {\sqrt {\b}}~ \partial_a \th~
 ~, ~~~~~~~~\w^{13}_a ~~\he~~  ~  {\sqrt {\b}}~ \sin \th~ \partial_a \p    \label{spin} 
  \eeq
  where $\b ~\equiv ~ {\frac {\a^2} {2e}}$ and  we have used $A(r_0) = {\frac {4r^2_0} e}$ with  $e  \equiv \exp (1) $.
 The corresponding curvature tensor components are:
\beq
 {\rm all}~~~R^{~~IJ}_{ab}(\w)~~~ \he~~0~ ~~{\rm except} ~~~ R^{~~23}_{ab}(\w)     
  ~~\he~~ 2 \sin \th~ \partial_{[a} \th ~ \partial^{}_{b]} \p 
  = {\frac {2 } { r^2_0 }}~ \S^{23}_{ab}~\equiv~ {\frac {2\g} { r^2_0 }}~ \S^{(1)}_{ab}
    \label{R}
 \eeq
 where we have introduced $\S^{(1)}_{ab}= \g^{-1}_{}\S^{23}_{ab}$.  These
 equations can be interpreted as a $U(1)$ Chern-Simons theory with $\w^{23}_a$ as the $U(1)$ gauge field.
    
 Notice, the connection component $\w^{~01}_a$ in (\ref{spin}) is  {\it pure
   gauge} and hence can be rotated away to zero by a boost gauge
 transformation $\w^{~IJ}_a  \rightarrow ~\w'^{IJ}_a$ where:
 \beq
  \w'^{01}_a &=& \w^{~01}_a- \partial_a \xi ~, ~~~~~~~~~~~~~~~~~~~~~~~~~ 
   \w'^{23}_a  =~  \w^{~23}_a\nonumber \\
   \w'^{02}_a & = &\cosh \xi~ \w^{~02}_a + \sinh \xi~ \w^{~12}_a ~,    
 ~~~~~~~ \w'^{03}_a  =~\cosh \xi~ \w^{~03}_a + \sinh \xi~ \w^{~13}_a ~,  \nonumber \\
  \w'^{12}_a & = &\sinh \xi~ \w^{~02}_a + \cosh \xi~ \w^{~12}_a ~,    
  ~~~~~~~\w'^{13}_a  =~\sinh\xi~ \w^{~03}_a + \cosh \xi~ \w^{~13}_a \nonumber
  \eeq
   From these,  if we choose $\xi = {\frac 1 2}\ln    ({\frac {2\b} c}) $ where $c$ is independent of the
  coordinates $v,~ \th, ~\p$, the connections fields   (\ref{spin}) then transform to:
  \beq
  \w'^{01}_a ~&\he&~ 0 ~~~~~~~~~~~~~~~~~~~~~ 
 \w'^{23}_a ~=~ -\cos \th~ \partial_a \p~, \nonumber \\
  \w'^{02}_a ~&\he&~  -~      {\frac {~c} {\sqrt 2}} ~ \partial_a \th~, 
 ~~~~~ \w'^{03}_a ~~\he~~ -~ {\frac {~c} {\sqrt 2}} ~  \sin \th  ~ \partial_a \p  \nonumber \\
 \w'^{12}_a ~& \he &       {\frac {~c} {\sqrt 2}} ~  \partial_a \th~
 ~, ~~~~~~~~~~\w'^{13}_a ~~\he~~      {\frac {~c} {\sqrt 2}} ~  \sin \th~ \partial_a \p   \label{spin'} 
  \eeq
 These are really  the gauge fields of $ISO(2)$ theory. To see this explicitly,
 we rewrite the fields as the following combinations:
  \beq
   {\cal A}^1_a &\equiv& \w'^{~23}_a ~~\he~~ -\cos \th ~\partial_a \p
~, \nonumber \\
{\cal A}_a^2   &\equiv &  {\frac 1 {\sqrt 2}}~ \left( \w'^{~02}_a -
  \w'^{~12}_a \right) ~~\he~~ - ~c ~ \partial_a \th~,   \nonumber \\
{\cal A}^3_a  &\equiv&  {\frac 1 {\sqrt 2}}~ \left( \w'^{~03}_a + \w'^{~31}_a \right)
~~\he~~ -~ c~ \sin \th~ \partial_a \p   \label{ISOA}
\eeq
 and
\beq
 {\tilde {\cal A}}^2_a   \equiv   {\frac 1 {\sqrt 2}}~ \left( \w'^{~02}_a + \w'^{~12}_a \right) ~\he~0~, 
 ~~~{\tilde {\cal A}}^3_a   \equiv   {\frac 1 {\sqrt 2}}~ \left( \w'^{~03}_a - \w'^{~31}_a \right)~\he~0 , ~~~ {\cal A}^4_a  \equiv \w'^{~01}_a ~\he~0
 \eeq
The  fields $({\cal A}^1_a,  ~ {\cal A}^2_a,~ {\cal A}^3_a) $ can be readily recognized as the 
three gauge fields of $ISO(2)$ subgroup. The three generators of
$ISO(2)  $ subgroup are given in terms of the generators of the Lorentz algebra $M_{IJ}$ by:
${\cal P}={\frac 1 {\sqrt 2}}\left(  K_2 - J_3 \right) ,$ $~{\cal Q}= {\frac 1 {\sqrt 2}}\left( K_3 + J_2\right)$,   and ${\cal  J}= J_1$.
where $ K_i ~\equiv~M_{0i}^{} = - M^{0i}_{}, ~~ J_i ~\equiv ~{\frac 1 2}~\epsilon_{ijk}~ M_{jk}$.
These satisfy the algebra: $ [{\cal P},~{\cal Q}] = 0, ~ [{\cal J},~{\cal P}]= Q,~  
[{\cal J},~ {\cal Q}]= -{\cal P} $.   This is the subgroup of Lorentz transformations that leave
a null internal vector invariant. For Schwarzschild spacetime, this null
vector is the Killing vector corresponding to the timelike
isometry of the exterior metric; on the horizon, this Killing vector turns
null. The $ISO(2)$ transformations correspond to the subgroup of local Lorentz
transformations which leave this vector invariant on the horizon \cite{bcg}. 

For the
$ISO(2)$ theory for the gauge fields (\ref{ISOA}), the  
field strength components satisfy the relations:
\beq
{\cal F}^1_{ab} &\equiv&  2~ \partial_{[a}^{} {\cal A}^1_{b]}  ~~\he~~ 2 \sin \th~ \partial_{[a}^{}\th ~\partial_{b]} \p ~ ~\he ~~ {\frac {2 \g}{  r^2_0}}~ \S^{(1)}_{ab} ~, \nonumber \\ 
{\cal F}^2_{ab} &\equiv&  2~   \partial_{[a}^{}   {\cal A}^2_{b]} ~+~ 
2~{\cal A}^1_{[a} {\cal A}^3_{b]}  ~~\he~~ 0~, \nonumber \\
 {\cal F}^3_{ab} &\equiv& 2~  \partial_{[a}^{}   {\cal A}^3_{b]} ~-~ 2~ {\cal A}^1_{[a} {\cal A}^2_{b]} ~~\he~~0 \label{ISOF}
\eeq
These equations are invariant under the $U(1)$ subgroup of   $ISO(2)$ 
gauge transformations. Hence these represent the equations of motion of 
an $ISO(2)$ Chern-Simons theory gauge fixed to $U(1)$ with source 
$\S^{(1)}_{\th \p}$  in the
direction of the $U(1)$ subgroup and coupling $k = {\frac {\pi r^2_0} {\g}}$.
To see that this indeed is the case consider the equations of motion of the $ISO(2)$ 
Chern-Simons theory  of gauge fields ${\cal A'}^i_a$ and their field strength ${\cal F}'^i_{ab}({\cal A}')$ with 
coupling constant $k$ and a specific source given by:
\beq
{\cal F}'^i_{v \th}({\cal A}')~ =~0~,   ~~~~~~~~~~{\cal F}'^i_{v\p}({\cal A}') ~=~0~,
~~~~~~~~~~{\frac k {2\pi}}~ {\cal F'}^i_{\th \p}({\cal A}') ~=~ J'^i_{} \label{ISOCS}
\eeq
These equations are covariant under the $ISO(2)$ gauge transformations which consist of two sets:
  (a) The $U(1)$ transformations, associated with the generators $ T_1 \equiv - {\cal J}$, on the gauge fields:  
\beq
   {\cal A }'^1_a  \rightarrow   {\cal A}'^1_a  - \partial_a^{} \a~,  
 ~~ {\cal A }'^2_a   \rightarrow     \cos \a~{\cal A}'^2_a  +  \sin \a~{\cal A}'^3_a~,   
 ~~{\cal A}'^3_a   \rightarrow      -~\sin \a ~ {\cal A}'^2_a  +  \cos \a~{\cal A}'^3_a
\label{ISOtransA1}
\eeq
where $\a$ is the local transformation parameter.
The field strength components change as:
\beq
   {\cal F}'^1_{ab}   \rightarrow    {\cal F}'^1_{ab}~,  
 ~~~ {\cal F}'^2_{ab}  \rightarrow    \cos \a~{\cal F}'^2_{ab} +\sin \a~ {\cal F}'^3_{ab}~,   
 ~~~{\cal F}'^3_{ab}  \rightarrow   -~\sin \a~{\cal F}'^2_{ab} + \cos \a~ {\cal F}'^3_{ab}
\label{ISOtransF1}
 \eeq 
  (b) The transformations associated with the generators $T_2\equiv -P,~~ T_3\equiv -Q$:
 \beq
  {\cal A' }^1_a \rightarrow    {\cal A'}^1_a   ~,  
 ~~~~{\cal A' }^2_a \rightarrow     {\cal A'}^2_a -  \partial_a^{}     c_2^{} -    {\cal A'}^1_a ~c_3^{}~,  
~~~~ {\cal A' }^3_a \rightarrow        {\cal A'}^3_a -  \partial_a^{}    c_3^{} +   {\cal A'}^1_a ~c_2^{} \label{ISOtransA2}
\eeq 
where $c_1$ and $c_2$ are two local  transformation parameters.
The field strength components   change as: 
\beq
   {\cal F' }^1_{ab} \rightarrow    {\cal F'}^1_{ab}~,
 ~~~~~ {\cal F' }^2_{ab}  \rightarrow      {\cal F'}^2_{ab}   -  {\cal F'}^1_{ab}~c^{}_3, 
~~~~~{\cal F' }^3_{ab}  \rightarrow     {\cal F'}^3_{ab}   +  {\cal F'}^1_{ab}~c^{}_2
\label{ISOtransF2}
 \eeq

Now, the first two equations of motion of   $ISO(2)$ Chern-Simons theory (\ref{ISOCS})
are satisfied by the   configurations where ${\cal A'}^i_v$ are pure gauge:
 \beq
 {\cal A}'^1_v &=&0~, ~~~~~~~~ {\cal A}'^2_v ~= ~-~\partial^{}_v c^{}_2~, ~~~~~~~~
 {\cal A}'^3_v ~=~ -~ \partial^{}_v c^{}_3~, \nonumber \\
 {\cal A}'^1_{\hat a}& =& {\cal B'}^1_{\hat a} ~, ~~~~~ {\cal A'}^2_{\hat a}~= ~
 {\cal B'}^2_{\hat a} -\partial^{}_{\hat a} c^{}_2 - {\cal B'}^1_{\hat a} c^{}_3~,
 ~~~~{\cal A}'^3_{\hat a}~= ~
 {\cal B'}^3_{\hat a} -\partial^{}_{\hat a} c^{}_3 + {\cal B'}^1_{\hat a} c^{}_2
 \eeq
 where ${\hat a} = (\th, \p)$ and ${\cal B'}^i_{\hat a}$ are independent of 
 the coordinate $v$. For these most general configurations $ {\cal F'}^i_{v \th}
 ({\cal A'}) =0$ and ${\cal F'}^i_{v \p}(
 {\cal A'}) =0$ hold identically and
 \beq
 {\cal F'}^1_{\th \p}({\cal A'})& = &{\cal F'}^1_{\th \p}({\cal B'}) ,
 ~~~~~~{\cal F'}^2_{\th \p} ({\cal A'})~=~
 {\cal F'}^2_{\th \p}({\cal B'}) - {\cal F'}^1_{\th \p}({\cal B'}) c^{}_3, \nonumber \\
  {\cal F'}^3_{\th \p} ({\cal A'}) &=& {\cal F'}^3_{\th \p}({\cal B'}) + {\cal F'}^1_{\th \p}(
 {\cal B'}) c^{}_2
 \eeq
>From (\ref{ISOCS}), the field strength components ${\cal F'}^i_{\th \p} ({\cal B'})$ satisfy the equations of motion:
\beq
{\frac k {2\pi}}~ {\cal F'}^i_{\th \p}({\cal B'})= {\tilde J}^i_{}~ ~~~{\rm where}~~ {\tilde J}^i_{} = J'^1_{}~,~~
{\tilde J}^2_{} = J'^2_{} + c^{}_3 J'^1_{}~,~~ {\tilde J}^3 = J'^3_{} - c^{}_2 J'^1_{}
\label{ISOCS1}
\eeq
For these equations, we are now left with invariance under $v$-independent 
$ISO(2)$ gauge transformations. We use this freedom to make a  gauge transformation  
of the type (b) above, by choosing the transformation parameters $c^{}_2(\th, \p)$ and $c^{}_3(\th, \p)$ appropriately: ${\cal B'}_{\hat a} \rightarrow {\cal A}^i_{\hat a}$~ and ~
${\cal F'}^i_{\th \p}({\cal B'})
\rightarrow {\cal F}^i_{\th \p}({\cal A })$ ~such that ~
${\cal F}^1_{\th \p}({\cal A}) \ne 0$,
 ~ ${\cal F}^2_{\th \p} ({\cal A} ) = 0$ ~and ~ ${\cal F}^3_{\th \p} ({\cal A} ) = 0.$
Consistent with this, the sources in (\ref{ISOCS1})  transform as: ${\tilde J}^i_{} \rightarrow J^i_{}$ where
 $J^i_{} = ( J, ~~0, ~~0)$ and  finally
  the equation of motion (\ref{ISOCS1}) lead to:
\beq 
{\frac k {2\pi}}~ {\cal F}^1_{\th \p}({\cal A}) = J~, ~~~~{\cal F}^2_{\th \p} ({\cal A} ) = 0~, ~~~~{\cal F}^3_{\th \p} ({\cal A} ) = 0 \label{ISOCS2}
\eeq
which, for $k= {\frac {\pi r^2_0}  {\g}}$ and $J= \S^{(1)}_{\th \p}$,  are same as   (\ref{ISOF}). These  equations are
    invariant under the left over $U(1)$ transformations 
(type (a)) above. Thus we have demonstrated that the equations (\ref{ISOF}) are a partially gauge fixed version 
of the $ISO(2)$ Chern-Simons equations (\ref{ISOCS}) with a left over invariance only under $U(1)$ transformations.

 \section{$SU(2)$ Chern-Simons Boundary Theory}
 Now we shall discuss that the horizon degrees of freedom can as well be described by a Chern-Simons theory of Barbero-Immirzi $SU(2)$ gauge fields.
 To see this, we notice that  the $SU(2)$  gauge fields (\ref{BIA}) on $\D$ are:
 \beq
  A^{(1)}_a &\he& {\frac {\g} 2}~ \partial_a \ln \b+\cos \th~\partial_a \p~, ~~~~ A^{(2)}_a ~\he~ - ~  {\sqrt {\b}}
  \left( \g ~\partial_a \th ~- ~\sin \th ~\partial_a \p \right)~, \nonumber \\
   A^{(3)}_a &\he& - ~ {\sqrt {\b}}
  \left( \g \sin \th ~ \partial_a \p~ + ~\partial_a \th \right) \label{A'}
  \eeq
   
 and the  field strength components   satisfy the following relations
  on $\D$:
  \beq
  F^{(1)}_{ab} &=& 2~\partial^{}_{[_a }A^{(1)}_{b]}  
  + 2~A^{(2)}_{[a} A^{(3)}_{b]}   ~~\he~~ -~{\frac 2 {r^2_0}}~ \left( 1 -  K^2_{} \right) ~\S^{23}_{ab} \nonumber \\ 
  F^{(2)}_{ab} &=& 2~\partial^{}_{[a} A^{(2)}_{b]}  
  +2~ A^{(3)}_{[a} A^{(1)}_{b]}  ~~\he~ ~  - ~2 {\sqrt {1+\g^2_{}}}~ \sin \th ~\partial_{[a} \p ~\partial_{b]} K \nonumber \\
  F^{(3)}_{ab} &=&2~ \partial^{}_{[a } A^{(3)}_{b]}  
  + 2~A^{(1)}_{[a} A^{(2)}_{b]}  ~~\he~~  2{\sqrt {1+\g^2_{}}}~  ~\partial^{}_{[a} \th ~\partial_{b]} K  
  \label{F'}
  \eeq
where $   K~ =~  {\sqrt { \b (1+\g^2_{})}}$ which is arbitrary through 
spacetime dependent   field $\b$ which can be changed by a boost transformation of the original tetrad and connection fields. Thus we may   gauge fix this invariance
under boost transformations   by a convenient choice of $\b$ as follows:

      (i) Now, $\b ~\equiv~ {\frac {\a^2_{}} {2e}}~=~ {\frac {-vw} {2e}} ~~\he~~ 0$ 
   ($K~\he~0$) is a possible 
  choice of the basis where the $SU(2)$ gauge fields from (\ref{A'}) are:
    \beq
  A^{(1)}_a ~~\he~~ {\frac {\g} 2}~ \partial^{}_a \ln v + \cos \th ~ \partial_a^{} \p~, ~~~~ 
  ~~~A^{(2)}_a~~\he~~0~, ~~~~~~~A^{(3)}_a ~~\he~~ 0   \label{A1}
  \eeq
  and from (\ref{F'}), the field strength components satisfy the following relations:
  \beq
  F^{(1)}_{ab} &=& 2~\partial^{}_{[_a }A^{(1)}_{b]}  
  + 2~A^{(2)}_{[a} A^{(3)}_{b]}   ~~\he~~ -~{\frac 2 {r^2_0}}~   ~\S^{23}_{ab} ~=~ -~{\frac {2\g} {r^2_0}}~   ~\S^{(1)}_{ab}
   \nonumber \\ 
  F^{(2)}_{ab} &=& 2~\partial^{}_{[a} A^{(2)}_{b]}  
  +2~ A^{(3)}_{[a} A^{(1)}_{b]}  ~~\he~~0  \nonumber \\
  F^{(3)}_{ab} &=&2~ \partial^{}_{[a } A^{(3)}_{b]}  
  + 2~A^{(1)}_{[a} A^{(2)}_{b]}  ~~\he~~  0
  \label{F1} 
\eeq 
Notice that these equations are unaltered under the $U(1)$
transformations: $A^{(1)}_a \rightarrow  A^{(1)}_a - \partial_a^{} \xi,$ ~
$A^{(2)}_a  \rightarrow \cos \xi A^{(2)}_a + \sin \xi A^{(3)}_a$ ~ and
$A^{(3)}_a \rightarrow  - \sin \xi A^{(2)}_a  + \cos \xi A^{(3)}_a$.  Hence,
these equations  can be interpreted as the equations of motion of a $SU(2)$
Chern-Simons theory gauge fixed to a $U(1)$ theory described by the $U(1)$
gauge field $A^{(1)}_a$ with coupling $ k = {\frac {\pi r^2_0} {\g}}$ and
source $ \S^{(1)}_{\th \p} \equiv {\g}^{-1}_{} \S^{23}_{\th \p}$ in the $U(1)$
direction. An important property to note is that here $\S^{(2)}_{\th \p}
~\equiv~ {\g}^{-1}_{}\S^{31}_{\th \p} ~\he~ 0 $ and $  \S^{(3)}_{\th
\p}\equiv~ {\g}^{-1}_{}\S^{12}_{\th \p} ~\he~ 0$.
  
  (ii) Other possible choice of the basis is where $\b$ is constant 
  (K constant), but arbitrary. Here the gauge fields are:
 \beq
     A^{(1)}_a &\he&  \cos \th~\partial_a \p~,
      ~~~~ A^{(2)}_a ~\he~ - ~  K 
  \left( \ \cos \d~\partial_a \th ~- ~\sin \d \sin \th ~\partial_a \p \right)~, \nonumber \\
   A^{(3)}_a &\he& - ~ K
  \left( \sin \d ~\partial_a \th ~+ ~\cos \d \sin \th ~ \partial_a \p~ \right)  \label{A2}
  \eeq 
  where   $K = {\sqrt {\b(1+\g^2_{})} } $ is now a constant and $\cot \d =  \g$.
    The right hand sides of last two equations in (\ref{F'}) are zero, that is, the field strength components satisfy:
  \beq
  F^{(1)}_{ab} &=& 2~\partial^{}_{[ a }A^{(1)}_{b]}  
  + 2~A^{(2)}_{[a} A^{(3)}_{b]}   ~~\he~~  
   -~{\frac {2\g} {r^2_0}}~ \left[ 1 - \b \left( 1 + \g^2\right) \right] ~\S^{(1)}_{ab}\nonumber \\ 
  F^{(2)}_{ab} &=& 2~\partial^{}_{[a} A^{(2)}_{b]}  
  +2~ A^{(3)}_{[a} A^{(1)}_{b]}  ~~\he~   ~0 \nonumber \\
  F^{(3)}_{ab} &=&2~ \partial^{}_{[a } A^{(3)}_{b]}  
  + 2~A^{(1)}_{[a} A^{(2)}_{b]}  ~~\he~~ 0
  \label{F2}
  \eeq
   In these equations, we may interpret 
     the combination
     \beq
  k = {\frac {\pi r^2_0} {\g}} ~\equiv ~ {\frac {a^{}_H} {4\g}}~,~~~~~ a^{}_H ~\equiv~ 4\pi r^2_0{}  \label{k}
  \eeq
  as the $SU(2)$ Chern-Simons coupling constant and the source as 
  \beq J^{(i)} = \left(\left[ 1 - \b\left( 1+\g^2\right) \right]   \S^{(1)}_{\th \p}~,~~~0~,~~~0 \right) \label{J}
  \eeq
  There is an arbitrary constant  parameter $\b$ in the source which can be
  changed by a boost transformation of the original spin connection fields.
  Notice that for $\b = \left( 1+\g^2\right)^{-1}$, the source vanishes. 
 
Alternatively,  we may take the combination
$ k =  {\frac {a_H^{}} {4 \g\left[ 1 - \b\left( 1+\g^2\right) \right]}}~   $
as the coupling constant of the $SU(2)$ Chern-Simons theory and $J^{(i)}=\left(\S^{(1)}_{\th \p}~,~0~,~0\right) $
as the source in the $U(1)$ direction of this theory. Then, we have a gauge
dependent arbitrariness in the  coupling constant, reflected through the
parameter $\b$. By boost transformations of the original spin connection fields, the value of
$\b$ can be changed. For the specific choice $\b = {\frac 1 2}$, we have the case of  
\cite{per}.  Also for $\b = \left( 1+\g^2\right)^{-1}$, the coupling constant
diverges. The ambiguity in how we define the Chern-Simons coupling strength
depends on how we define bulk sources for the horizon Chern-Simons theory,
which in turn depends on our choice of Lorentz frame in the bulk used to define
the Schwarzschild spacetime in terms of tetrad frame components.   
  
 Like the  equations of motion (\ref{F1}),
  eqns.  (\ref{F2}) have invariance under the left over $U(1)$ transformations.
  Thus, the gauge theory described by Eqns. (\ref{A1})-(\ref{J}), can be viewed as  
  a  $SU(2)$ Chern-Simons theory on the horizon $ \D $ with a specific set of sources partially gauge fixed to $U(1)$.  
  To see this explicitly, consider the $SU(2)$ Chern-Simons theory with coupling $k$ described by the action:
 \beq 
S_{CS}~= ~{\frac k { 4\pi}} \int_{\D }^{}
\e^{abc}_{} \left( A'^{(i)}_a \partial^{}_b A'^{(i)}_c ~+ ~ {\frac  1 3}
~\e^{ijk}_{}  A'^{(i)}_a A'^{(j)}_b A'^{(k)}_c  \right)   ~+~ \int_{\D  }
J'^{(i)a}_{} A'^{(i)}_a ~~ \label{CSAction+} 
\eeq 
Here the nonzero components of
the completely antisymmetric $\e^{abc}_{}$ are given by $ \e^{v \theta \p}_{}= 1$ 
and  the   source, which is a vector density with upper index
$a$,  is covariantly conserved, $ D_a(A') J'^{(i)a}_{} $ $\equiv~
\partial^{}_a J'^{(i)a}_{} + \e^{ijk}A'^{(j)}_a J'^{(k)a}_{} =0$, and further
has the special form as: 
\beq J'^{(i)a}_{} ~\equiv~ \left( J'^{(i)v}_{},~
J'^{(i)\theta}_{}, ~J'^{(i)\p}_{} \right) ~=~ \left( J'^{(i)}_{}, ~~0~, ~~0
\right) \label{sources} \eeq
  The action
(\ref{CSAction+}) is independent of the metric of the three-manifold $\D$.
  
  Now the equations of motion for the $SU(2)$ Chern-Simons action (\ref{CSAction+}) are:
\beq
 F'^{(i)}_{v \theta}(A')  ~~\he~~0~, ~~~~~~~~~F'^{(i)}_{v \p}(A') ~~\he~~0  ~,  ~~~~~~~~~
 {\frac k {2\pi}  }~F'^{(i)}_{\theta \p}(A')   ~~\he~~ - ~  J'^{(i)}_{}   \label{CSKruskal1}
\eeq
The most general solution of the first two equations in this set is provided by  the configurations where
$A'^{(i)}_v$ are pure gauge:
\beq
A'^{(i)}_v = -{\frac 1 2}~ \e^{ijk}_{} \left( {\cal O} \partial_v^{} {\cal O}^T_{} \right)^{jk}_{},
~~~~ A'^{(i)}_ {\hat a} = {\cal O}^{ij}_{} B'^{(j)}_{\hat a} - 
{\frac 1 2}~ \e^{ijk}_{} \left( {\cal O} \partial_{\hat a}^{} {\cal O}^T_{} \right)^{jk}_{},
~~~~{\hat  a} = (\th, ~\p) \label{sol}
\eeq
with the $ SU(2)$ gauge   fields $ B'^{(i)}_\th $  and $B'^{(i)}_\p$ independent of
$v$. Here ${\cal O}$ is an arbitrary $3 \times 3$ orthogonal matrix, 
${\cal O} {\cal O}^T_{} = {\cal O}^T_{} {\cal O} =1$ with $det~ {\cal O} =1$. As $F'^{(i)}_{v \th}
(A') ~~ \he ~~0$ and $F'^{(i)}_{v \p} (A') ~~\he~~ 0$ are  identically satisfied, from (\ref{CSKruskal1}),
we are left with the equation:
\beq
{\frac k {2\pi}}~ F'^{(i)}_{\th \p} (A') ~=~ {\frac k {2\pi}}~  {\cal O}^{ij}_{} F'^{(j)}_{\th \p}(B')  
 ~ ~\he ~~ -~   J'^{(i)} \label{CSB'}
\eeq
 where $F'^{(i)}_{\th \p}(B')$ is the $SU(2)$ field strength for the 
 gauge fields ~$(B'^{(i)}_\th,~ B'^{(i)}_\p )$.  
 
 This solution (\ref{sol}) has fixed part of the   $SU(2)$ gauge invariance;
for the fields  $B'^{(i)}_\th(\th, \p)$ and $  B'^{(i)}_\p(\th, \p)$, we are
now left with invariance only  under  $v-$independent $SU(2)$ gauge transformations
on the spatial slice $S^2$ of $\D$.  Using this freedom, through a $v-$independent  
 transformation  matrix $  {\bar {\cal O}}(\th, \p)$,
 it is always possible
to write the triplet of  field strength  $F'^{(i)}_{\th \p}(B')$ in terms
of a field strength which is parallel to a unit vector $u^i_{}(\th ,\p)$ in the internal space:
\beq 
F'^{(i)}_{\th
\p}(B') &=& {\bar  {\cal O}}^{ij}_{} F^{(j)}_{\th \p} (B) ~\equiv~ u^i(\th, \p)~ F^{(1)}_{\th \p}(B)~, 
\nonumber \\
     F^{(1)}_{\th \p} (B) &\ne &0~, ~~~~~ F^{(2)}_{\th \p} (B)~= ~0,  ~~~~~
F^{(3)}_{\th \p} (B)~=~0 \label{FB} 
\eeq 
where $u^i_{}(\th,\p) ~\equiv~ {\bar {\cal O}}^{i1}(\th, \p) $ and the
gauge fields $B'^{(i)}_{\hat a}(\th, \p)$ and  $B^{(i)}_{\hat a}(\th, \p)$
with the index $ {\hat a} = (\th ~, \p)$  are related by a gauge transformation as:
\beq 
B'^{(i)}_{ \hat a} = {\bar {\cal O}}^{ij}_{}
B^{(i)}_{\hat a } - {\frac 1 2}~ \e^{ijk} \left( {\bar  {\cal O}} \partial_{\hat a}^{}
{\bar {\cal O}}\right)^{jk}_{}\label{B}
\eeq
 
As discussed in the Appendix, there are two types of gauge fields
$B^{(i)}_{\hat a}(\th,\p)$ that  yield the field strength, as in (\ref{FB}),
parallel to the unit vector $ u^i_{} $    which we may parameterized as
$u^i_{} (\th, \p) = \left( \cos \Theta , ~\sin \Theta \cos \Phi , ~\sin \Theta
\sin \Phi\right)$ in terms of two angles $\Theta (\th, \p) $ and $\Phi(\th,
\p)$. These two types are, from (\ref{U1B2}) and (\ref{U1B3}): (i) $
B^{(i)}_{\hat a} =\left( B^{}_{\hat a} +\cos \Theta~ \partial_{\hat a}^{}\Phi,
~0, ~0\right)$ with $B^{}_{\hat a}$ as arbitrary. This corresponds to the
configuration (\ref{A1})  with its field strength as in (\ref{F1}) above for
$B^{}_{\hat a}=0$ and $\Theta = \th, ~ \Phi = \p$.  (ii) The second solution
is: $ B^{(1)}_{\hat a} =- \partial_{\hat a}^{} \d+ \cos \Theta~ \partial_{\hat
a}^{} \Phi$,  $ B^{(2)}_{\hat a} = c \left( \cos \d ~\partial_{\hat a}^{}
\Theta - \sin \d  \sin \Theta ~\partial^{}_{\hat a}\Phi\right)$, $
B^{(3)}_{\hat a} =   c \left( \sin \d ~\partial_{\hat a}^{} \Theta + \cos \d
\sin \Theta ~\partial^{}_{\hat a}\Phi\right)$ where $c$ is a constant and  $\d
( \th, \p)$ is   arbitrary. Notice that this configuration is the same as
that describing   the horizon fields in (\ref{A2}) with $c=-K$ and $\Theta \th, ~ \Phi=\p $ and $\d$ as constant. The corresponding field strength
components satisfy the equations of motion given by  (\ref{F2}) with the
coupling   and the sources as identified by (\ref{k}) and (\ref{J}).
 
Thus we may rewrite the equations (\ref{sol}), for both these cases,   as:
\beq A'^{(i)}_v = -{\frac 1 2}~ \e^{ijk}_{} \left( {\cal O'} \partial_v^{}
{\cal O'}^T_{} \right)^{jk}_{}, ~~~ A'^{(i)}_ {\hat a} = {\cal O'}^{ij}_{}
B^{(j)}_{\hat a} -  {\frac 1 2}~ \e^{ijk}_{} \left( {\cal O'} \partial_{\hat
a}^{} {\cal O'}^T_{} \right)^{jk}_{} \label{sol1} \eeq where ${\cal O'} {\cal O} {\bar {\cal O}}$. The field strength components $F'^{(i)}_{v \th}(A')
$ and $F'^{(i)}_{v \p} (A')$ are identically zero and  the equation
(\ref{CSB'}) becomes \beq {\frac k {2 \pi}  }~ F'^{(i)}_{\th \p} (A') ~=~
{\frac k {2 \pi} }~ {\cal O'}^{ij}_{} F^{(j)}_{\th \p}(B)   ~ ~\he ~~ -~
J'^{(i)} ~ \equiv ~-~  {\cal O'}^{ij} J^{(j)} ~  \label{CSB} \eeq where now
from (\ref{FB}),  $F^{(i)}_{\th \p}(B) = \left( F^{(1)}_{\th \p} (B), ~0,~0
\right)$, which implies for  the sources 
\beq 
J^{(i)} = \left( J, ~~0,~~0 \right)  \label{sourcesU1} 
\eeq 
As discussed in the Appendix, in terms of the
fields   $B^{(i)}_{\hat a}$ and the corresponding field strength $ \left(
F^{(1)}_{\th \p} (B), ~0,~0 \right)$, we have a theory with left over
invariance only under $U(1)$ gauge transformations.  Thus, in the $SU(2)$
theory partially gauge fixed to a theory with  invariance only under $U(1)$
transformations, {\it the sources in the internal space directions orthogonal
to the $U(1)$ are zero; the only source is in  the direction of the $U(1)$
subgroup.} Further, the coupling constant of the $SU(2)$ Chern-Simons theory
is given by $k = {\frac {\pi r^2_0} {\g}}$.

\section{Concluding remarks}

That there are different, but equivalent, classical formulations of the topological theory of the 
horizon degrees of freedom is to do with the fact that it is essentially only  the properties 
of various fields on the spatial slice $S^2_{}$ of the horizon that are
relevant. Note in this respect that our approach is quite complementary to the
Hamiltonian analysis of isolated horizons \cite{ash1}, \cite{per}.
Though  our analysis here has been
restricted to the case of the event horizon of the Schwarzschild solution, 
many of our conclusions do 
in fact generalize for generic spherical isolated horizons. However,
a Hamiltonian analysis of the constraints of the theory in
presence of isolated horizons described by a set of boundary conditions, as
has been done in the quoted references, could be performed. 
Classical Hamiltonian formulation of
the $SU(2)$ Chern-Simons theory on the event horizon
has three first class constraints corresponding to the 
three generators of $SU(2)$ gauge transformations. On three-manifolds 
with  topology of $S^2\times R$, in the process of  gauge fixing 
from $SU(2)$ to $U(1)$, two of these are gauge fixed through gauge fixing
constraints with which these form a set of second class constraints
according to the standard rules of gauge fixing. To implement these
second class constraints, we need to go over from the Poisson brackets
to the corresponding Dirac brackets. We are then left with only one
first class constraint associated with the left over $U(1)$ invariance.

In Loop Quantum Gravity where the bulk properties are described
by the quantum theory based on Barbero-Immirzi $SU(2)$ gauge theory, the
horizon degrees of freedom are described by  an $SU(2)$ Chern-Simons theory,
or equivalently its gauge fixed version  in terms of  a theory exhibiting 
only a left over $U(1)$ invariance but
with additional constraints on the solder forms. Further, 
there are no local degrees of freedom in topological quantum Chern-Simons theories; 
all the degrees of freedom are global or topological. These global degrees 
of freedom reside
in the properties of the punctures on $S^2$. These are given by the spin networks 
from the bulk quantum theory where we have $SU(2)$ spins living on these punctures. This information 
is in the values of the solder forms on $S^2$ which, in the quantum theory, have distributional 
support at these punctures. 

Properties of the
quantum black holes can be calculated in  either formulation, $SU(2)$ or the 
partially gauge fixed version  with only the left over 
$U(1)$ invariance, yielding the
same results. In particular, the black hole entropy in either formulation has
the standard leading area law and  the subleading correction given by
logarithm of area with  definite coefficient $-3/2$ for large black hole area
as obtained in \cite{kmprl, dkmprd}.  The value of the Barbero-Immirzi parameter $\g$
obtained by matching the leading area term with the Bekenstein-Hawking law is
also the same.  However, as already mentioned and also emphasized earlier in \cite{bkm}, 
care needs to be exercised in doing the calculations
in the $U(1)$ formulation  by implementing the extra conditions on the solder
forms on the quantum states contributing to the entropy.
  
Though, in the realistic situation of a sufficiently massive star collapsing
gravitationally, the past horizon ($v=0$) of the idealized Kruskal-Szekeres extended 
Schwarzschild geometry is never realized, it is of
interest  to note that the discussion developed above holds for this horizon
also. Its degrees of  freedom are again  described by an $SU(2)$ Chern-Simons
theory, or equivalently its gauge fixed version in terms of a $U(1)$ theory.
 
  \acknowledgments 
  Discussions with Rudranil Basu, Ayan Chatterjee and
Amit Ghosh are gratefully  acknowledged. RKK thanks
Ghanashyam Date for his comments and  Lee Smolin for correspondence regarding
early work on  Chern-Simons theory  in the context of non-perturbative quantum gravity
and also  acknowledges the support of the Department of Science and Technology, 
Government of India through  a J.C. Bose Fellowship.
 
\appendix* 
 
\section{ $U(1)$ Gauge theory  as gauge fixed $SU(2)$ Chern-Simons  on $S^2$}
 
Here we explicitly discuss how the $SU(2)$   fields $B'^{(i)}_{\hat
a}(\th,\p)$ of Eqn. (\ref{B}) can  be gauge fixed to the fields $B^{(i)}_{\hat
a}$ with only a left over $U(1)$ invariance on the spatial slice $S^2$ of the
horizon.  To unravel the nature of the  fields  $B^{(i)}_{\hat a}(\th, \p)$,
we may parametrize these  along the internal space unit vector $u^i_{} (\th,
\p)  $ and orthogonal to it as: \beq B'^{(i)}_{ \hat a } ~=~   u^i~ B_{\hat a}
~+~ f ~\partial_{\hat a} u^i  ~+~ g ~\e^{ijk} u^j \partial_{\hat a} u^k~,
~~~~~~ {\hat a} = ( \th,~\p) \label{U1A}   \eeq  where $f$ and $g$ are
functions on the spacetime $S^2$. Then the six independent field degrees  of
freedom in     $B'^{(i)}_{ \hat a}$ are now distributed in $u^i$ (two
independent fields), $B_{\hat a}$ (two field degrees of freedom) and the two
fields ($f,~ g$). Had the internal space unit vector $u^i$ in (\ref{U1A}) been
completely arbitrary, all these six field degrees of freedom would be
independent. Here   $u^i$ is, through (\ref{FB}), parallel to the field
strength $F'^{(i)}_{\th \p}(B') $.  For this reason, all of these six field
degrees of freedom are not independent.
  
Now the field strength for the gauge fields (\ref{U1A}) can be calculated
in a straight forward manner to be: \beq F'^{(i)}_{{\hat a}{\hat b}}(B') &=&
u^i \left( 2~\partial_{[{\hat a}} B_{{\hat b}]} + \left( f^2 + g^2 +2g \right)
\epsilon^{ijk} u^j \partial_{\hat a} u^k \partial_{\hat b} u^l \right)
\nonumber \\ &+& 2~\partial_{[{\hat a}} u^i\left( (1+g) B_{{\hat b}]}
- \partial_{{\hat b}]} f \right) - 2~\epsilon^{ijk} u^j\partial_{[{\hat a}}
u^k \left( f B_{{\hat b}]} + \partial_{{\hat b}]} g\right) \label{FB1} \eeq
where we have used the identity for the internal space unit vector $u^i_{}$:
\begin{equation} \label{id1} \epsilon^{ijk}  \partial_{\hat a}
u^j \partial_{\hat b} u^k  ~=~ u^i ~\epsilon^{jkl} u^j \partial_{\hat a}
u^k \partial_{\hat b} u^l
\end{equation}
Now comparing (\ref{FB1}) with (\ref {FB}) where the components orthogonal to $u^i_{}$ are zero, we
have the conditions:
 \beq
\partial_{\hat a} f - (1+g)B_{\hat a} =0~, ~~~~~~~~ \partial_{\hat a} g + fB_{\hat a} =0 \label{cond}
\eeq
 These conditions imply that
   there are two classes of the $SU(2)$ gauge fields $B'^{(i)}_{ \hat a}$ 
  on the spatial slice $S^2_{}$ which give such a field strength: 
  
(i) The first kind are where the field $B^{}_{\hat a}$ is arbitrary and
$f~=~0$ and $1~+~g~=~0$. Then the gauge field      from (\ref{U1A}) is: \beq
B'^{(i)}_{ \hat a} ~=~ u^i~ B^{}_{\hat a} ~-~ \e^{ijk}~ u^j_{}\partial_{\hat
a}^{} u^k_{}
  \label{U1A'} \eeq Clearly the field strength for such a gauge field is
parallel to $u^i_{}$: \beq F'^{(i)}_{{\hat a}  {\hat b}}(B') ~=~ u^i\left (2
~\partial^{}_{[{\hat a}}  B^{}_{{\hat b}]} - \e^{jkl}_{} u^j \partial_{\hat a}
u^k_{}\partial^{}_{\hat b} u^l_{} \right) \label{U1F'} \eeq  The quantity $
\epsilon^{{\hat a} {\hat b}}\epsilon^{jkl}_{} u^j_{} \partial_{\hat a}^{}
u^k_{} \partial_{\hat b}^{} u^l_{}$  is the {\it winding number}  density for
the homotopy maps $S^2 \rightarrow S^2$   and its integral over the
two-dimensional space $S^2$ is characterized by integers (Homotopy group
$\Pi_2(S^2) = \mathbb{Z}$).  Since it is a topological density we can write it
as  \beq \epsilon^{jkl}_{} u^j_{} \partial_{\hat a}^{} u^k_{} \partial_{\hat
b}^{} u^l_{}~=~ 2~\partial_{[\hat a}^{} \Omega_{\hat b]}^{} \label{Omega} \eeq
In particular, for the parameterization of  unit vector $u^i_{} $ in terms two
angles $  \Theta(\th, \p)$ and $\Phi(\th, \p)$ as  $ u^i_{} ~= ~ \left( \cos
\Theta, ~ \sin \Theta \cos  \Phi , ~ \sin \Theta \sin \Phi \right) $,  we
have,  \beq  \epsilon^{jkl}_{} u^j_{} \partial_{\hat a}^{}
u^k_{} \partial_{\hat b}^{} u^l_{} ~=~    2~\sin \Theta  ~\partial_{[\hat
a}^{} \Theta~   \partial_{{\hat b}]}^{} \Phi~ ~~~~~~~ {\rm  and }~~
\Omega_{\hat a}^{}= -  \cos \Theta ~
\partial^{}_{\hat a} \Phi.
\eeq
Further,   for the unimodular orthogonal matrix ${\bar {\cal O}}^{ij}_{}$
with its components as:  $ {\cal {\bar O}}^{i1}_{} = u^i_{} =   \left( \cos
\Theta, ~ \sin \Theta \cos  \Phi , ~ \sin \Theta \sin \Phi \right)$ , ${\cal
{\bar O}}^{i2}_{} = \left( - \sin \Theta, ~ \cos \Theta \cos  \Phi , ~ \cos
\Theta \sin \Phi \right) $,  ~ $ {\cal {\bar O}}^{i3}_{} =  \left( ~0~, ~ -
\sin  \Phi , ~ \cos   \Phi \right)$, the following identity can be shown to
hold: \beq \e^{ijk}_{} u^j_{} \partial^{}_{\hat a} u^k_{} ~=~ u^i_{}
~\Omega_{\hat a} ~+~ \frac 1 2~ \e^{ijk}_{} {\bar {\cal
O}}^{jl}_{} \partial^{}_{\hat a} {\bar {\cal O}}^{kl}_{}  \label{id2} \eeq
Defining ${\cal B}_{\hat a}   ~\equiv~ B^{}_{\hat a} - \Omega^{}_{\hat a}$,
this identity allows us to rewrite the gauge field  (\ref{U1A'}) and its field
strength (\ref{U1F'})   as: \beq B'^{(i)}_{ \hat a} &=& u^i~ {\cal B}^{}_{\hat
a}~ - {\frac 1 2}~ \e^{ijk} \left( {\bar  {\cal O}}
 \partial^{}_{\hat a } {\bar {\cal O}}^T \right)^{jk}_{}  ~\equiv~ {\bar {\cal O}}^{ij}_{}
 B^{(j)}_{ \hat a} -     {\frac 1 2}~ \e^{ijk} \left( {\bar  {\cal O}}
 \partial^{}_{\hat a } {\bar {\cal O}}^T \right)^{jk}_{}   , \nonumber \\
  F'^{(i)}_{ {\hat a}  {\hat b }}(B') &=& 2~u^i~ \partial^{}_{[{\hat a}} {\cal  B}^{}_{{\hat b}]}
  ~\equiv ~2~{\bar {\cal O}}^{ij}_{} ~\partial^{~}_{[{\hat a}}  B^{(j)}_{{\hat b}]}
  \label{U1B2} \eeq with the $SU(2)$ gauge field $ B^{(i)}_{ \hat a}~\equiv~
\left( B^{(1)}_{ \hat a}, ~B^{(2)}_{ \hat a}, ~B^{(3)}_{ \hat a} \right) ~=~
\left( {\cal B}_{\hat a}^{}, ~~0,~~0 \right)$.  Thus,   the  fields ${\cal
B}^{}_{\hat a}$ and their field strength ${\cal F}^{}_{{\hat a} {\hat b}}
=2 \partial^{}_{[{\hat a}} {\cal  B}^{}_{{\hat b}]}$,  describe  $U(1)$  gauge
configurations obtained from    $SU(2)$  gauge fields  by partial gauge
fixing.
  
(ii) The second class of gauge fields are where $B_{\hat a}^{}  ~= ~-
~\partial^{}_{\hat a} \d$, ~$f =  c  \cos \d$ and $1 + g =  c  \sin \d$  with
$c$ as a constant and $\d(\th, \p)$ an arbitrary field. For these values we
have, $f^2+(1+g)^2=c^2$. Thus, the gauge fields  from (\ref{U1A}) are: \beq
B'^{(i)}_{ \hat a} ~=~ - u^i~\partial^{}_{\hat a} \d + c\cos \d
~ \partial^{}_{\hat a} u^i_{} +  \left( c \sin \d - 1\right)~ \e^{ijk}_{}
u^j_{} \partial_{\hat a}^{} u^k_{} \label{U1A''} \eeq For these gauge fields,
the field strength is again parallel to $u^i_{} ( \th, \p)$: \beq
F'^{(i)}_{{\hat a}  {\hat b}}(B') ~=~\left( c^2_{} -1 \right)~ u^i  ~
\e^{jkl}_{} u^j \partial_{\hat a} u^k_{}\partial^{}_{\hat b} u^l_{}  \eeq
which can again be rewritten as: \beq F'^{(i)}_{{\hat a}  {\hat b}}(B') ~=~
u^i ~{\cal F}^{}_{{\hat a} {\hat b}}~,  ~~~~~ {\cal F}^{}_{{\hat a} {\hat b}}
~ = ~2~\partial^{}_{[{\hat a}} {\cal  B}^{}_{{\hat b}]} ~ ~~ ~~{\rm
with}~~~~~{\cal  B}^{}_{\hat a} ~  = ~c^2_{} \Omega_{\hat a}
~-~\Omega^{}_{\hat a} \label{U1F''} \eeq where $\Omega^{}_{\hat a}$ is given
by (\ref{Omega}). This   field strength then is completely characterized by
the $U(1)$ theory of gauge field ${\cal B}^{}_{\hat a}$ and its field strength
${\cal F}^{}_{{\hat a}{\hat b}}$.
   
Using the parametrization for the unit vector $u^i$ as earlier and the
identity (\ref{id2}), the gauge fields  (\ref{U1A''})  can be rewritten as:
\beq B'^{(i)}_{ \hat a} ~=~  {\bar {\cal O}}^{ij} ~ B^{(j)}_{\hat a} - {\frac
12}~\e^{ijk} {\bar {\cal O}}^{jl} \partial^{}_{\hat a} {\bar {\cal
O}}^{kl}~\label{U1B22} \eeq where, in this case the gauge field $B^{(i)}_{\hat
a}$ has all   three internal space components non-zero, but of a specific
form: \beq B^{(1)}_{\hat a} &=&~    -~ \partial^{}_{\hat a}\d ~+~\cos
\Theta~\partial^{}_{\hat a} \Phi~, \nonumber \\ B^{(2)}_{\hat a} &=&  ~c
~\left( \cos \d~\partial^{}_{\hat a} \Theta~-~\sin \d \sin \Theta
~\partial^{}_{\hat a} \Phi \right), \nonumber \\ B^{(3)}_{\hat a} &=&  c~
\left( \sin \d~\partial^{}_{\hat a} \Theta~+~ \cos \d \sin
\Theta~ \partial^{}_{\hat a} \Phi \right) \label{U1B3} \eeq This partially
gauge fixed $SU(2)$   theory described by the gauge fields  (\ref{U1B2}) and
(\ref{U1B3}) with  field strength (\ref{U1F''}), has only a left over gauge
invariance   under   $U(1)$ transformations: $ B^{(1)}_{\hat a}  \rightarrow
B^{(1)}_{\hat a} +  \partial^{}_{\hat a} \l$,~ $~ B^{(2)}_{\hat a} \rightarrow
\cos \l~B^{(2)}_{\hat a} + \sin \l~B^{(3)}_{\hat a}$, ~ $ B^{(2)}_{\hat a}
\rightarrow  - \sin \l~B^{(2)}_{\hat a} + \cos \l~B^{(3)}_{\hat a}$. We may
use this invariance to rotate away the arbitrary field $\d$ to zero in
(\ref{U1B3}) through an $U(1)$ transformation with $\l = \d$.
  
An interesting property to note is  that  the gauge configurations
(\ref{U1A'}) of class (i) satisfy the condition that the unit vector $u^i$ is
covariantly constant, that is, $D^{}_{\hat a}(B') u^i
\equiv  \partial^{}_{\hat a} u^i_{} + \e^{ijk}_{} B'^{(j)}_{\hat a} u^k_{}  0$. On the other hand, the  configurations corresponding to the class (ii)
given by   (\ref{U1A''})   do not satisfy this condition; instead these
satisfy: $D^{}_{\hat a}(B') u^i =  c\left(\sin\d~ \partial^{}_{\hat a}u^i_{} -
\cos \d~ \e^{ijk}_{}    u^j \partial_{\hat a} u^k \right) \ne 0$ for $c \ne
0$.
   
For $c=0$, the configuration (\ref{U1B3}) of class (ii)   reduces to
(\ref{U1A'}) of the class (i) above with $B^{}_{\hat a} =0$. But for
$c~\ne~0$, the configurations of  these two classes are  completely distinct.

In fact there is a general  underlying  mathematical reason for the fact
that $SU(2)$ Chern-Simons theory on a manifold $R\times S^2$ with punctures on
$S^2$, like the horizon, can be described by a $U(1)$ theory with consequent
conditions on the special sources (\ref{sources}) of the $SU(2)$ Chern-Simons
theory that these are zero in the directions orthogonal to   $U(1)$ direction
given by the internal space vector $u^i$ as given in (\ref{sourcesU1}). This
is as follows: In general, the gauge group $SU(2)$, with its group manifold
being  $S^3$, can be  mapped to a  $S^2$ ($SU(2)/U(1)$ coset space) in  such a
way that every point on $S^2$ comes from  a circle on $S^3$. Such maps, known
as Hopf maps, generate the homotopy group $\Pi_3(S^2)$ which is just the set
of integers $\mathbb{Z}$. So $SU(2)$ is a bundle  (the Hopf bundle) over $S^2$
with a $U(1)$ fibre. Now, any $SU(2)$ gauge field $(B'^{(i)}_\th,
B'^{(i)}_\p)$ in two dimensional spacetime $S^2$ can be, in general, gauge
fixed in the internal space directions contained in the coset space
$SU(2)/U(1)$ $\cong S^2 $ (which are orthogonal to  the internal space unit
vector  $u^i$ characterizing  this $S^2$). This  gauge fixing  can be achieved
through an appropriate  gauge transformation  from the coset space
$SU(2)/U(1)$.   Then we are just left  with   invariance under the   $U(1)$
transformations which leave this unit  vector $u^i$ unaltered. Thus an
arbitrary  $SU(2)$ gauge  field $(B'^{(i)}_\th , B'^{(i)}_\p)$ on spacetime
$S^2$ is completely characterized by the internal space unit vector $u^i$  and
the  gauge field $({\cal B}_\th, {\cal B}_\p)$  of the $U(1)$ subgroup in the
direction $u^i$.   The corresponding $SU(2)$ field strength is   parallel to
$u^i$ and is  completely determined by  the $U(1)$ field strength.

\end{document}